\documentclass[12pt,a4paper]{article}

\usepackage{graphicx}
\usepackage{amsmath}
\usepackage{amssymb}
\usepackage[font=small]{caption}
\usepackage{fullpage}
\usepackage[numbers]{natbib}
\usepackage{color}
\usepackage[pagewise]{lineno}
\usepackage{soul}
\usepackage{ulem}

\usepackage{enumerate}
\usepackage[titletoc,title]{appendix}

\usepackage{setspace}

\usepackage[T1]{fontenc}
\usepackage[utf8]{inputenc}
\usepackage{authblk}
\usepackage{multirow}
\title{Ultrasensitivity on signaling cascades revisited: Linking local and global ultrasensitivity estimations }
\author[1]{Edgar Altszyler}
\author[2,3]{Alejandra C. Ventura}
\author[3,4]{Alejandro Colman-Lerner}
\author[5,6]{Ariel Chernomoretz \thanks{Corresponding author: ariel@df.uba.ar}}
\affil[1]{Laboratorio de Inteligencia Artificial Aplicada, Universidad de Buenos Aires, Departamento de Computaci\'on, Ciudad Universitaria, 1428 Buenos Aires, Argentina - CONICET}
\affil[2]{Universidad de Buenos Aires, Facultad de Ciencias Exactas y Naturales.}
\affil[3]{CONICET-Universidad de Buenos Aires, Instituto de Fisiología, Biología Molecular y Neurociencias (IFIBYNE), Buenos Aires, Argentina.}
\affil[4]{Universidad de Buenos Aires, Facultad de Ciencias Exactas y Naturales, Departamento de Fisiología, Biología Molecular y Celular.}
\affil[5]{Departamento de F\'isica FCEN UBA - IFIBA CONICET.}
\affil[6]{Fundación Instituto Leloir.}

\date{\vspace{-5ex}}
\begin{document} 
\maketitle


\begin{abstract}
Ultrasensitive response motifs, which are capable of converting graded stimulus in binary responses, are very well-conserved in signal transduction networks.
Although it has been shown that a cascade arrangement of multiple ultrasensitive modules can produce an enhancement of the system’s ultrasensitivity, how the combination of layers affects the cascade’s ultrasensitivity remains an open question for the general case. Here we introduced a methodology that allowed us  to determine the presence of sequestration effects and to quantify the relative contribution of each module to the overall cascade’s ultrasensitivity. The proposed analysis framework provides a natural link between global and local ultrasensitivity descriptors and is particularly well-suited to characterize and better understand mathematical models used to study real biological systems. As a case study we considered three mathematical models introduced by O’Shaughnessy et al. to study a tunable synthetic MAPK cascade, and showed how our methodology might help modelers to better understand modeling alternatives. 

\end{abstract}


\section{Introduction}

Sigmoidal input-output response modules are very well-conserved in cell signaling networks. They might be used to implement binary responses, a key element in cellular decision processes. Additionally, sigmoidal modules might be part of more complex structures, where they can provide the nonlinearities which are needed in a broad spectrum of biological processes \cite{ferrell2014ultrasensitivity3,zhang2013ultrasensitive}, such as multistability \cite{angeli2004detection,ferrell2001bistability}, adaptation \cite{srividhya2011open}, and oscillations \cite{kholodenko2000negative}. There are several molecular mechanisms that are able to produce sigmoidal responses such as inhibition by a titration process \cite{buchler2008molecular,buchler2009protein}, zero-order ultrasensitivity in covalent cycles \cite{goldbeter1981amplified,ferrell2014ultrasensitivity}, and multistep activation processes - like multisite phosphorylation \cite{ferrell2014ultrasensitivity2,ferrell1996tripping, ferrell1997responses, markevich2004signaling,gunawardena2005multisite} or ligand binding to multimeric receptors \cite{rippe1997analysis}.

Sigmoidal curves are characterized by a sharp transition from low to high output following a slight change of the input. The steepness of this transition is called ultrasensitivity \cite{ferrell2014ultrasensitivity}. In general, the following operational definition of the Hill coefficient may be used to calculate the overall ultrasensitivity of sigmoidal modules:

\begin{equation}
 n_H = \frac{\log(81)}{\log(EC90/EC10)} \label{hill}
\end{equation}

where EC10 and EC90 are the signal values needed to produce an output of 10\% and 90\% of the maximal response, respectively. The Hill coefficient $n_H$ quantifies the steepness of a transfer function relative to the hyperbolic response function which is defined as not ultrasensitive and has $n_H=1$. This value means that an 81-fold increase in the input signal is required to change the output level from $10\%$ to $90\%$ of its maximal value. Response functions with $n_H>1$ need a smaller input fold increase to produce such output change, and are thus called ultrasensitive functions. 

Global sensitivity measures such the one described by equation \ref{hill} do not fully characterize s-shaped curves, y(x), because they average out local characteristics of the analyzed response functions. Instead, these local features are well captured by the logarithmic gain or response coefficient measure \cite{kholodenko1997quantification} defined as:

\begin{equation}
 R(x) = \frac{x}{y} \frac{dy}{dx} = \frac{d\log(y)}{d\log(x)} \label{R_def}
\end{equation}

Equation \ref{R_def} provides local ultrasensitivity estimates given by the local polynomial order of the response function.

\subsubsection*{MAP kinase cascades}

Mitogen activated protein (MAP) kinase cascades are a well-conserved motif. They can be found in a broad variety of cell fate decision systems involving processes such as proliferation, differentiation, survival, development, stress response and apoptosis \cite{keshet2010map}. They are composed of a chain of three kinases which sequentially activate one another, through single or multiple phosphorylation events. A thoughtful experimental and mathematical study of this kind of systems was performed by Ferrell and collaborators, who analyzed the steady-state response of a MAPK cascade that operates during the maturation process in Xenopus oocytes \cite{huang1996ultrasensitivity}. They developed a biochemical model to study the ultrasensitivity displayed along the cascade levels and reported that the combination of the different ultrasensitive layers in a multilayer structure produced an enhancement of the overall system's global ultrasensitivity \cite{huang1996ultrasensitivity}. 
In the same line, Brown et al. \cite{brown1997protein} showed that if the dose-response curve of a cascade, $F(x)$, could be described as the mathematical composition of functions, $f^{is}$, each of which describe the behavior of each layer in isolation (i.e,  $F(x)=f^{is}_{MK}(f^{is}_{MKK}(f^{is}_{MKKK}(x)) )$, then the local ultrasensitivity of the different layers combines multiplicatively:  $R(x)=R_{MK}(f^{is}_{MKK}(f^{is}_{MKKK}(x)) . R_{MKK}(f^{is}_{MKKK}(x) ) .  R_{MKKK}(x)$. In connection with this result, Ferrell showed, for the special case of two Hill-type modules of the form

\begin{equation}
 y = k\frac{x^{n_H}}{{EC50}^{n_H}+x^{n_H}} \label{hill_function}
\end{equation}

(where the parameter EC50 corresponds to the value of input that elicits half-maximal output, and $n_H$ is the Hill coefficient), that the overall cascade global ultrasensitivity had to be less than or equal to the product of the global ultrasensitivity estimators of each cascade's layer, i.e $n_H \leq n_{H,1} \,  n_{H,2}$ \cite{ferrell1997responses}. 

Hill functions of the form given by equation \ref{hill_function} are normally used as empirical approximations of sigmoidal dose-response curves, even without any mechanistic foundation \cite{zhang2013ultrasensitive}. However, it is worth noting that for different and more specific sigmoidal transfer functions, qualitatively different results could have been obtained. In particular, a supra-multiplicative behavior (the ultrasensitivity of the combination of layers is higher than the product of individual ultrasensitivities) might be observed for left-ultrasensitive response functions, i.e. functions that are steeper to the left of the EC50 than to the right. In this case, the boost in the ultrasensitivity is caused by  the asymmetrical dose-response functional form (see \cite{altszyler2014impact} for details).

As modules are embedded in larger networks, constraints in each module's input's dynamic range could arise. We formalized this idea in a recent publication introducing the notion of dynamic range constraint of a module's dose-response function. The later concept is a feature inherently linked to the coupling of modules in a multilayer architecture, and resulted a relevant element to explain the overall ultrasensitivity displayed by a cascade \cite{altszyler2014impact}. Besides dynamic range constraint effects, sequestration (i.e. the reduction in free active enzyme due to its accumulation in complex with its substrate) is another relevant process inherent to cascading that could reduce the cascade's ultrasensitivity \cite{bluthgen2006effects,racz2008sensitivity,wang2016tunable}. Moreover, sequestration may alter the qualitative features of any well-characterized module when integrated with upstream and downstream components, thereby limiting the validity of module-based descriptions \cite{ventura2008hidden,del2008modular,ventura2010signaling}. 

All these considerations expose the relevance of studying the behavior of modular processing units embedded in large networks. Although there has been significant progress in the understanding of kinase cascades, how the combination of layers affects the cascade's ultrasensitivity remains an open question for the general case. 

In the present work, we have developed a method to describe the overall ultrasensitivity of a kinase cascade in terms of the effective contribution of each module. We used our approach to analyze a recently presented synthetic MAPK cascade experimentally engineered by O'Shaughnessy et al. \cite{o2011tunable}. 

Using a synthetic biology approach O'Shaughnessy et al. \cite{o2011tunable} constructed an isolated mammalian MAPK cascade (a Raf-MEK-ERK system) in yeast and analyzed its information processing capabilities under different rather well-controlled environmental conditions. They made use of a mechanistic mathematical description to account for their experimental observations. Their model was very similar in spirit to Huang-Ferrell's \cite{huang1996ultrasensitivity} with two important differences: a) no phosphatases were included, and b) the creation and degradation of all species was explicitly taken into account. Interestingly, they reported that the multilayer structure of the analyzed cascades can accumulate ultrasensitivity supra-multiplicatively, and suggested that cascading itself and not any other process (such as multi-step phosphorylation, or zero-order ultrasensitivity) was at the origin of the observed ultrasensitivity. They called this mechanism, de-novo ultrasensitivity generation. As we found the proposed mechanism a rather appealing and 
unexpected way of ultrasensitivity generation, we wanted to further characterize it within our analysis framework. In particular, we reasoned that the methodology and concepts introduced in the present contribution were particularly well-suited to understand the mechanisms laying behind the ultrasensitivity behavior displayed by O'Shaughnessy cascade model. 

The paper is organized as follows. First, we present a formal connection between local and global descriptors of a module's ultrasensitivity for the case of a cascade composed of $N$ units. We then introduce the notion of Hill input's working range in order to analyze how a module embedded in a cascade contributes to the overall system's ultrasensitivity.  Next, we describe a simple methodology to identify the presence of sequestration effects that might affect the system ultrasensitive behavior. Finally, as a case study, we present the O'Shaughnessy cascade analysis in order to show the insights that might be gained using the introduced concepts and analysis methodologies. We conclude by presenting a summarizing discussion and the conclusions of the work.

\section*{Results}
\subsection*{Linking local and global ultrasensitivity estimations}

The concept of ultrasensitivity describes a module's ability to amplify small changes in input values into larger changes in output values. It is customary to quantify and characterize the extent of the amplification both globally, using the Hill coefficient $n_H$ defined in equation \ref{hill}, and locally, using the response coefficient, R(I), as a function of the module's input signal I (equation \ref{R_def}), 
We found a simple relationship between both descriptions considering the logarithmic amplification coefficient $A^f_{a,b}$, defined as:

\begin{equation}
 A^f_{a,b}=\frac{\log(f(b))-\log(f(a))}{\log(b)-\log(a)} \label{amplification}
\end{equation}

\begin{figure}[h]
    \begin{center}
    \includegraphics[width=1\textwidth]{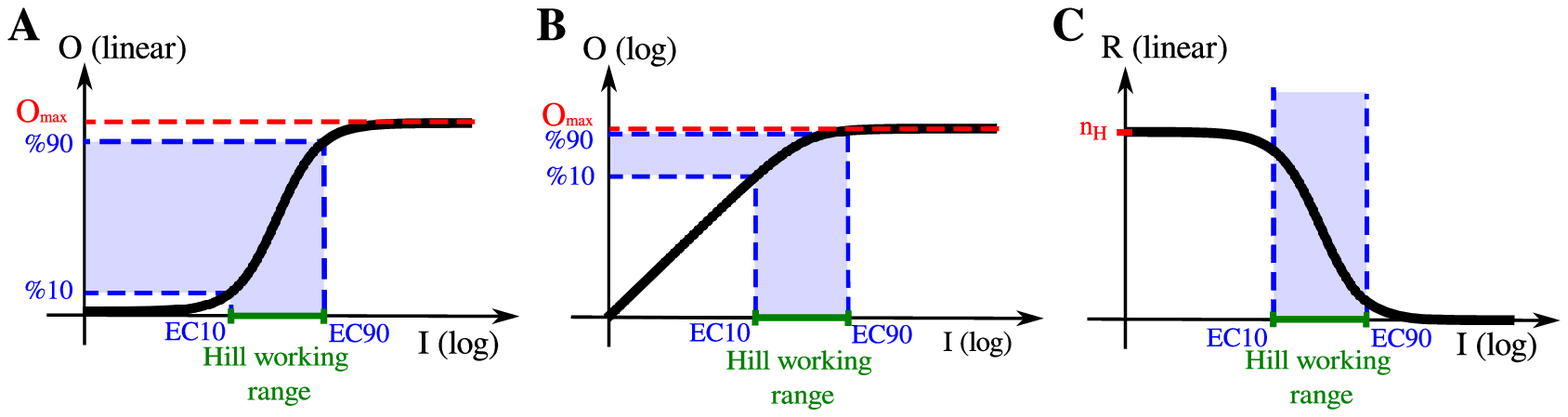}
    \caption{{\bf Hill function dose-response.}\\
    Schematic representations of a Hill type dose-response curve are shown in log-linear scale (A) and in log-log scale  (B). The EC10 and EC90 input levels are the signal values needed to produce an output level of 10\% and 90\% of the maximal response ($O_{max}$). The \textit{Hill input’s working range}, HIWR,  is the input range relevant for the calculation of the system’s $n_{H}$. For isolated modules HIWR = [EC10,EC90]. Panel (C) displays the corresponding local ultrasensitivity curve. Note that for Hill functions, inputs much smaller than the EC50 value present local sensitivities levels around their Hill coefficient value.} \label{fig_hill_dr}.
    \end{center}
\end{figure}

$A^f_{a,b}$ describes  the change (in a logarithmic scale) produced in the output when the input varies from a to b values. For instance,  $A^f_{a,b}= 0.5$ for an hyperbolic function evaluated between the inputs that resulted in 90\% and 10\% of the maximal output. In this case, the two considered input levels delimited the input range that should be considered for the estimation of the respective Hill coefficient $n_H$. We call this input interval: the  \textit{Hill input’s working range} (HIWR) (see Fig \ref{fig_hill_dr}A-B). 

Taking into account equation \ref{amplification}, the parameter $n_H$ can be rewritten as follows,
\begin{equation}
 n_H = \frac{\log(81)}{\log(EC90/EC10)} = \frac{2 \log(0.9/0.1)}{\log(EC90/EC10)}=2 A^f_{EC10,EC90} = \frac{A^f_{EC10,EC90}}{A^{hyp}_{EC10,EC90}} \label{nh_vs_amplification}
\end{equation}

Consequently, the Hill coefficient could be interpreted as the ratio of the logarithmic amplification coefficients of the function of interest and an hyperbolic function, evaluated in the corresponding \textit{Hill input’s working range}. 

It is worth noting that the logarithmic amplification coefficient that appeared in equation \ref{nh_vs_amplification} equaled the slope of the line that passed through the points $(EC10, f(EC10))$ and $(EC90, f(EC90))$ in a log-log scale. Thus, this quantity equals the average response coefficient calculated over the interval $HIWR=[EC10, EC90]$ in logarithmic scale (see Fig \ref{fig_hill_dr}C). Therefore,

\begin{equation}
 n_H  =  2 A^f_{EC10,EC90} =2 \frac{\int^{\log(EC90)}_{\log(EC10)} R_f(I) d(\log I)}{ \log(EC90)-\log(EC10)} =  2 \langle R_f \rangle_{EC10,EC90} = 
 \frac{\langle R_f \rangle_{EC10,EC90}}{\langle R_{hyp} \rangle_{EC10,EC90}} \label{nh_vs_R}
\end{equation}

where $\langle X \rangle_{a,b}$ denoted the mean value of the variable x over the range [a,b]. 

This last equation explicitly links the local and global ultrasensitivity descriptions. In particular, it can be appreciated that the module’s Hill coefficient is the average response coefficient over the module’s \textit{Hill input’s working range}, in units of a reference hyperbolic curve.

\subsection*{Ultrasensitivity in function composition. }
We generalize the last result to cast the overall global ultrasensitivity level of a multitier cascade in terms of logarithmic amplification coefficients. We proceed by first considering two coupled ultrasensitive modules, disregarding effects of sequestration of molecular components between layers. In this case, the expression for the system’s dose-response curve, $F$, results from the mathematical composition of the functions, $f_i$, which describe the input/output relationship of isolated modules $i=1,2$:

\begin{equation}
F(I_1)=f_2\big(f_1(I_1)\big) \label{F_composition}
\end{equation}

\begin{figure}[!h]
    \begin{center}
    \includegraphics[width=1\textwidth]{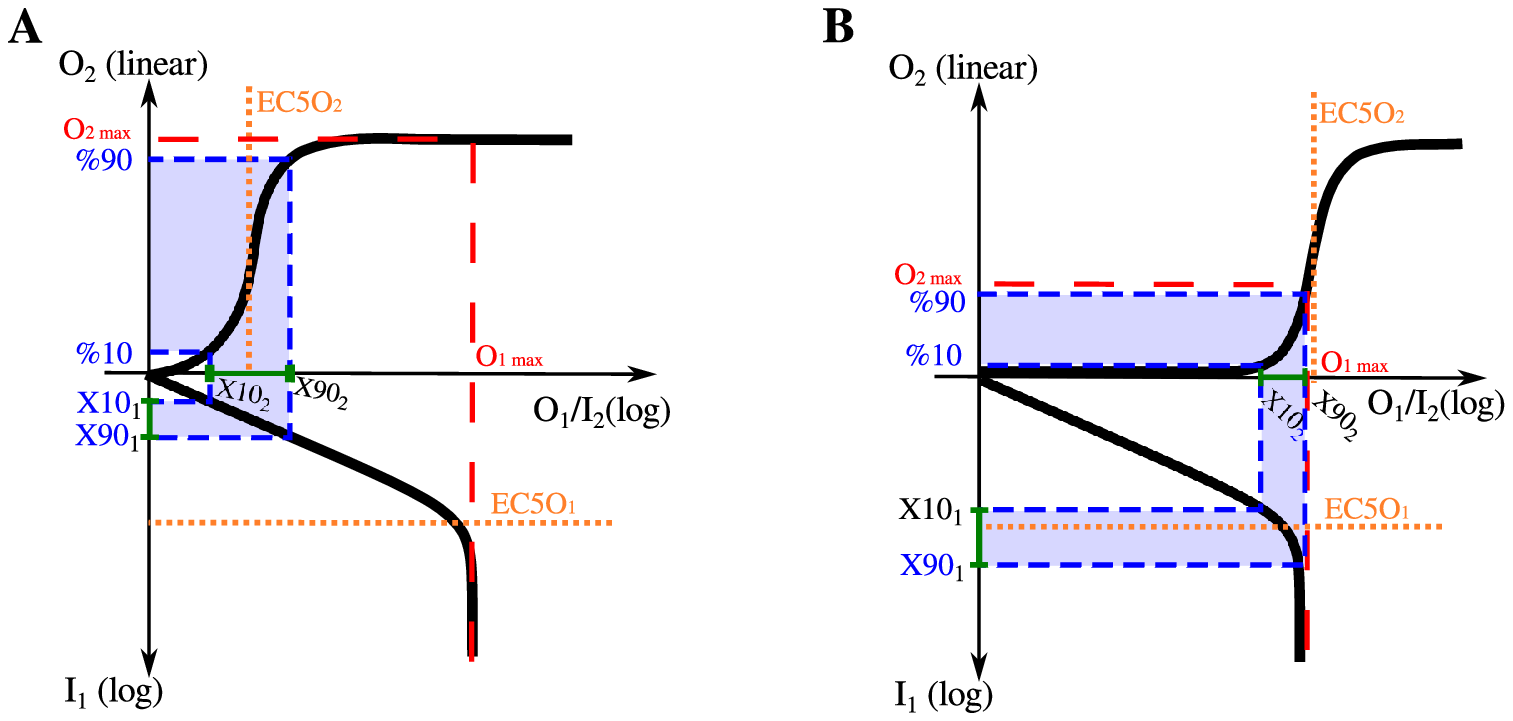}
    \caption{{\bf Hill functions composition.}\\
    Schematic response function diagrams for two different compositions of two Hill type ultrasensitive modules.  In each panel, the dose-response function of the first module is displayed in the lower semi-plane: the downward vertical axis representing the first module's input signal while its response function (that corresponds to the second module's input) is displayed along the horizontal axis. The dose-response curve for the second module is displayed in the upper-plane.  In panel (A) the maximum output level of the first module largely exceeds the EC50 level of the second one: $O_{1,max} \gg EC50_2$, whereas $O_{1,max} < EC50_2$ in panel (B).}
   
    \label{fig_hill_composition}
    \end{center}
\end{figure}

Using equation \ref{nh_vs_amplification}, the system’s Hill coefficient $n_H$ can be written as:

\begin{eqnarray}
 n_H &=& \frac{\log(81)}{\log(X90_1/X10_1)} =2 \overbrace{\frac{\log(0.9/0.1)}{\log(X90_2/X10_2)}}^{\nu_2}\overbrace{\frac{\log(X90_2/X10_2)}{\log(X90_1/X10_1)}}^{\nu_1} \nonumber \\
&=&2 \overbrace{A^{f_2}_{X10_2,X90_2}}^{\nu_2} \overbrace{A^{f_1}_{X10_1,X90_1}}^{\nu_1} = 
2 \overbrace{\langle R_2 \rangle_{X10_2,X90_2}}^{\nu_2} \overbrace{\langle R_1 \rangle_{X10_1,X90_1}}^{\nu_1} \label{decomposition_2_layers} \\
&=& 2 \,\nu_2 \, \nu_1  \nonumber 
\end{eqnarray} 

where $X10_i$ and $X90_i$ are the boundaries of the \textit{Hill input’s working range} of the composite system, i.e. the input values for the i-layer that produce a 10\% and 90\% of the system's maximal response , respectively (see Fig \ref{fig_hill_composition}).

Fig. \ref{fig_hill_composition} shows schematic response function diagrams for two different compositions of two Hill type ultrasensitive modules. Panel (A) corresponds to a situation where the maximum output level of the first module largely exceeds the EC50 level of the second one: $O_{1,max}\gg EC50_2$.
In this case $O_{2,max}$ equals the maximum output level of module 2 in isolation, $X10_2$ and $X90_2$ match the EC10 and EC90 levels of module 2 in isolation and the \textit{Hill input’s working range} of module 1 is located in the input region below $EC50_1$.
Panel (B) on the other hand, shows a setup where $O_{1,max} < EC50_2$. In this case $O_{2,max}$ is less than the maximum output level of module 2 in isolation, and the HIWR [$X10_2$, $X90_2$] differ from the input range [EC10,EC90] that should have been considered for module 2 in isolation. As a result, module-1's HIWR is centered at values higher than the corresponding EC50 level.\\

It follows from equation \ref{decomposition_2_layers} that the system’s Hill coefficient $n_H$ depends on the product of two factors, $\nu_1$  and $\nu_2$, which characterized local average sensitivities over the relevant input region for each layer: $[X10_i,X90_i ]$, with $i=1,2$ (see Fig \ref{fig_hill_composition}). We call the $\nu_i$ coefficient: effective response coefficient of layer-i.

For the more general case of a cascade of $N$ modules we found that:

\begin{equation}
 n_H = 2 \overbrace{ \langle R_N \rangle_{X10_N,X90_N}}^{\nu_N} \overbrace{\langle R_{N-1} \rangle_{X10_{N-1},X90_{N-1}}}^{\nu_{N-1}} .... \overbrace{\langle R_1 \rangle_{X10_1,X90_1}}^{\nu_1}= 2 \, \nu_{N} \, \nu_{N-1}...\nu_{1} \label{nh_vs_mu}
\end{equation}


This last equation shows a very general result. For the general case, the overall $n_H$ of a cascade could be understood as a multiplicative combination of the $\nu_i$ of each module. In this way, the effective response coefficients allow us to characterize the relative contribution of each layer to the overall system’s ultrasensitivity. 

It is worth noting that the factor two in Eq.\ref{nh_vs_mu} comes from the average response coefficient of a reference hyperbolic curve that appears in the original definition of the Hill coefficient (see Eq.\ref{nh_vs_R}). Hence, formally, the ultrasensitivity character of the cascade remains a system level feature, as it involves the product of the effective coefficient of all layers, in units of the logarithmic amplification coefficient of a reference hyperbolic curve.

\subsection*{The effect of the Hill’s input working range in multi-tiered systems.} 

According to equation (\ref{nh_vs_mu}) the \textit{Hill’s input working range} of a module bounds the relevant region of inputs over which local-ultrasensitivity features of module's response functions are combined to build up the overall system behavior.
It is thus a significant parameter to get insights about the overall ultrasensitivity of multilayered structures. In this section we show, for different types of dose-response curves, how this relevant interval depends on the way cascade layers are actually coupled. 

\subsubsection*{Composition of Hill functions}
Let’s start by considering two coupled ultrasensitive modules of the Hill type. Two different regimes can be identified depending whether the upstream module’s maximum output is large enough to fully activate the downstream unit (see Fig. \ref{fig_hill_composition})

\subsubsection*{Downstream saturation regime:}

In the first case i.e. when $O_{1,max} \gg  EC50_2$ (panel \ref{fig_hill_composition}A), $X10_2$ and $X90_2$ are equal to the respective $EC10_2$ and $EC90_2$ levels. Therefore, when coupled to module-1, the HIWR of module-2 does not differ from the one corresponding to the isolated case, and thus $\nu_2=\langle R_2 \rangle_{X10_2,X90_2} = n_2^{is}/2$.  In the last expression $n_2^{is}$ refers  to the Hill coefficient of module-2 when considered in isolation. On the other hand the HIWR of module-1 tends to be located at low input-values for increasing levels of the ratio $O_{1,max}/EC50_2$. In this region the response coefficient of the Hill functions achieve the highest values,  $R_1 \approx n_1^{is}$ (see Fig \ref{fig_hill_dr}C), thus, when calculating the average logarithmic gain, we  obtain $\nu_1 = \langle R_{1} \rangle_{X10_1,X90_1}=n_1^{is}$. Finally, following equation \ref{nh_vs_mu} we get
 \[
 n_H=2.\nu_1.\nu_2= 2.n_1^{is}.n_2^{is}/2=n_1^{is}.n_2^{is}. 
 \]
 It can be seen that the cascade behaves multiplicatively in this regime, which is consistent with Ferrell’s results \cite{ferrell1997responses}\\
 
\subsubsection*{Upstream saturation regime:}

When the upstream module’s maximal output does not fully activate the downstream module, i.e. $O_{1,max} \lesssim EC50_2$, different behaviors could arise depending on module-2 ultrasensitivity features at low input values:

\begin{figure}[h]
    \begin{center}
    \includegraphics[width=0.5\textwidth]{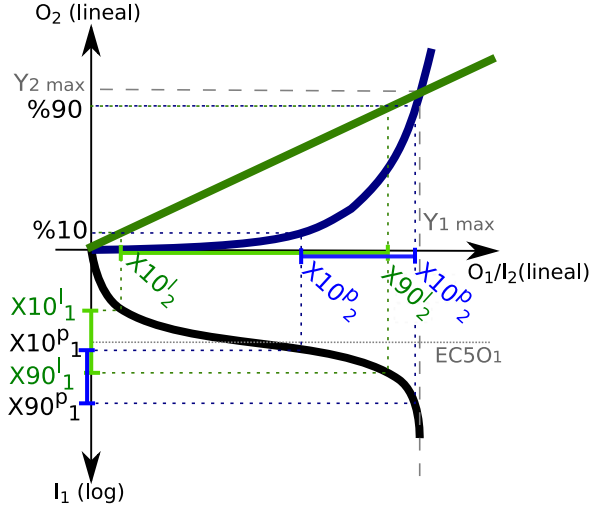}
    \caption{ Schematic response function diagrams for the composition of a Hill function as module-1, with a linear function (in green) or a power function (in blue) as module-2.}
    \label{fig_hills_composition}
    \end{center}
\end{figure}

For instance, let's consider that module-2 dose-response has $n_2=1$. This means that it displays a linear behavior at low input values (see Fig \ref{fig_hills_composition}). This linearity produces that $X10^l_2$ and $X90^l_2$  ($X10_2$ and $X90_2$ of the linear curve) match the \%10 and \%90 of  $O_{1, max}$. Therefore, $X10^l_1$ and $X90^l_1$ equals  $EC10_1$ and $EC90_1$ respectively, centering the HIWR around the $EC50_1$level. 
Furthermore, as a result of the linearity displayed by module-2 response function in this regime, the system’s overall behavior relies exclusively on module-1's ultrasensitivity and,  given the linearity of module-2, shows a multiplicative behavior. Applying equation \ref{nh_vs_mu}

\begin{equation*}
 n_H = 2\overbrace{ \langle R_2 \rangle_{X10_2,X90_2}}^{\nu_2} \overbrace{\langle R_1 \rangle_{X10_1,X90_1}}^{\nu_1} = 2 \overbrace{1}^{\nu_2} \overbrace{n_1/2}^{\nu_1}=n_1 
\end{equation*}

On the other hand, when $n_2>1$, module-2 dose response presents a power-law behavior for low input values (see Fig \ref{fig_hills_composition}). In this case, the non-linearity produces a shift in module’s-2 working range toward higher values, which centers module’s-1 the HIWR $[X10_1,X90_1]$  around input values higher than $EC50_1$.  Furthermore, given that $R_1$ decreases with $I_1$ (see Fig \ref{fig_hill_dr}C), the shift in module’s-1 working range results in $\nu_1 = \langle R_1 \rangle_{X10_1,X90_1} < n_1 /2$, and consequently,

\begin{equation*}
 n_H = 2\overbrace{ \langle R_2 \rangle_{X10_2,X90_2}}^{\nu_2} \overbrace{\langle R_1 \rangle_{X10_1,X90_1}}^{\nu_1} < 2 \overbrace{ n_2}^{ \nu_2} \overbrace{n_1/2}^{\nu_1}=n_2 n_1 
\end{equation*}
Therefore, whenever $n_2>1$, we get for upstream saturation ($O_{1,max} < EC50_2$) and $n_2 > 1$ cases , that  the system displays a submultiplicative behavior (consistent with Ferrell’s results \cite{ferrell1997responses}).

\subsubsection*{Golbeter-Koshland functions composition}

The detailed functional form of the response curve of an ultrasensitive module could deeply affect the overall system’s ultrasensitivity in cascade architectures.  In particular, a system composed by two modules characterized by Golbeter-Koshland, GK, response functions \cite{goldbeter1981amplified}, instead of a Hill type functional form, shows a qualitatively different behavior.

 \begin{figure}[h]
     \begin{center}
     \includegraphics[width=1\textwidth]{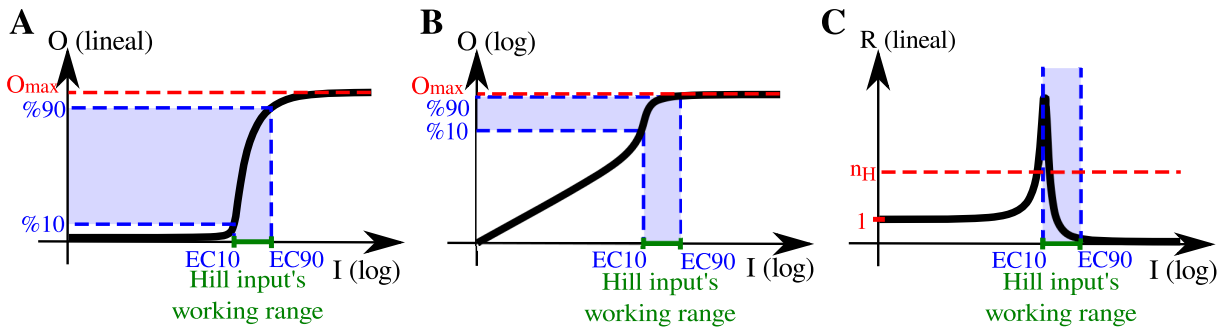}
     \caption{  Schematic representations of a Goldbeter-Koshland dose-response curve (for $K_1 \gtrsim 1$ and $K_2 \ll 1$, see equation in appendix \ref{S1_text} ) are shown in log-linear scale (A) and in log-log scale  (B).  The corresponding response coefficient, displayed in Panel (C), shows no local ultrasensitivity for small input values (i.e. $R\sim 1$) , but displays  high local sensitivity levels, even larger than the module's Hill coefficient $n_H$, for intermediate input regions.}\label{fig_GK}
     \end{center}
 \end{figure}

\begin{figure}[h]
    \begin{center}
    \includegraphics[width=1\textwidth]{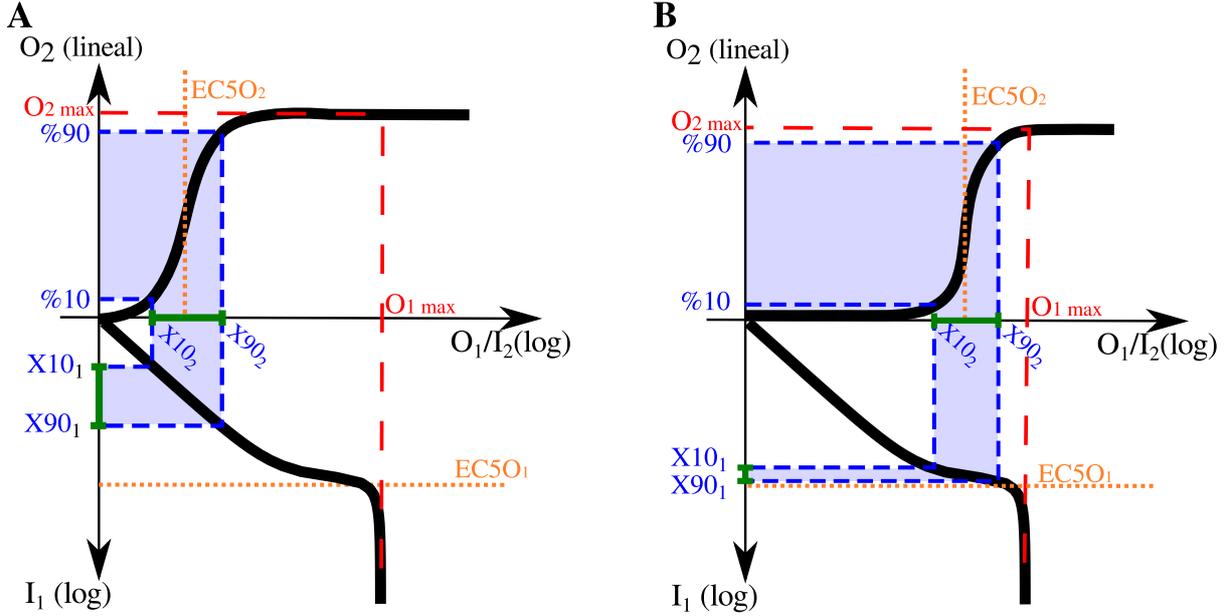}
    \caption{
    Schematic response function diagrams for two different compositions of two GK ultrasensitive modules are shown in panels (A) and (B).  Axes were arranged as explained in Fig.\ref{fig_hill_composition}'s caption. In panel (A)  $O_{1,max} \gg EC50_2$, and module-1's HIWR  lays in the input region below $EC50_1$ where the curve shows no local ultrasensitivity ($R_1=1$). In panel (B) we show a special setup where the $O_{2,max}/EC50_2$ ratio was tuned in order to set module-1's working range in its most ultrasensitive region.}
    \label{fig_GK2}
    \end{center}
\end{figure}

GK functions appear in the mathematical characterization of covalent modification cycles (such as phosphorylation-dephosphorylation), ubiquitous in cell signaling, operating in saturation (see appendix \ref{S1_text}). The detailed functional form of the transfer function depends on the operating regimes of the phosphorylation and dephosphorylation processes \cite{gomez2007operating}. For cases where the phosphatases, but not the kinases, work in saturation, GK functions present input regions with response coefficients higher than their overall $n_H$ \cite{altszyler2014impact}  (see Fig \ref{fig_GK}A-C). Hence, whenever their HIWR is located in the region of largest local ultrasensitivities, these functions are able to contribute with more effective ultrasensitivity than their global ultrasensitivity value. This means that cascades involving GK functions may exhibit supra-multiplicative behavior.

Fig. \ref{fig_GK2}A shows that for a two tier arrangement of these kind of modules under downstream saturation regime (i.e. when module’s-2 EC50 is much lower than the GK maximal output level, $O_{1,max}$) the module’s-1 HIWR is set in its linear response regime (i.e. $R_1=1$), and the GK function does not contribute to the overall system’s ultrasensitivity. However, a particular  $O_{1,max}/EC50_2$ ratio value exists for which module’s-1 HIWR spans the most ultrasensitive region of the module's transfer function, producing an effective response coefficient, $\nu_2$, even larger than the overall ultrasensitivity of the GK curve in isolation (i.e. $\nu_2 \geq n_2$ ), resulting in supra-multiplicative behavior $n_H=2. \nu_1.\nu_2 >n_1^{is}. n_2^{is}$ (see Fig \ref{fig_GK2}B ).

Comparing the Hill and GK cases, our analysis highlights the impact of the detailed functional form of a module’s response curve on the overall system’s ultrasensitivity in cascade architectures. Local sensitivity features of the involved transfer functions are of the utmost importance in this kind of setting and could be at the core of non-trivial phenomenology

\subsection*{Disentangling the contribution of HIWR and sequestration effects on observed ultrasensitivities}

As we have shown in the preceeding sections, the resetting of HIWRs induced by module coupling could be at the core of the system's ultrasensitivity. In addition, sequestration effects, affecting free active enzyme concentrations due to intermediary complex formation, could also play an important role at this respect \cite{bluthgen2006effects,racz2008sensitivity,wang2016tunable}. Sequestration and dynamic range constraints not only contribute with their individual complexity, but also usually occur together, thus making it more difficult to identify their individual effective contribution to the system's overall ultrasensitivity.

In order to disclose the impact of these two factors we simultaneously consider two approximations of the system under study (see Fig. \ref{S1_Fig} in appendix).
For a given model, we first consider the mathematical compositions of each module's response function (e.g. for a MAPK cascade $ {F}_{MAPK}^{non-seq} (x) = {f}_{MAPK}^{is} \Big({f}_{MAPKK}^{is}\big({f}_{MAPKKK}^{is}(x)\big)\Big)$, see \ref{S1_Fig}B in appendix). We name this expression ${F} ^ {non-seq}$ because, by construction, sequestration effects are completely neglected in this transfer function. On the second hand we also numerically estimate the response function $F^{seq}$, obtained by numerical integration of the the corresponding mechanistic model of the cascade \ref{S1_Fig}C in appendix) . 

In this way, the first estimation, $F^{non-seq}$, allows us to analyze to what extent the existence of HIWRs impinges on ultrasensitivity features of the cascade arrangement of layers. On the other hand, $F^{seq}$ not only incorporates HIWR resetting effects, but also serves to assess for putative sequestration effects that could take place in the system (see \ref{S1_Fig} in appendix).

\subsection*{Ultrasensitivity in O’Shaughnessy \textit{et al.} models}

In this section we aim to revisit three different mathematical models proposed by O’Shaughnessy \textit{et al}.  to disentangle the origin of the ultrasensitive behavior observed in a mammalian MAPK cascade engineered in yeast \cite{o2011tunable}. In particular, we will show how the methodology and concepts introduced so far can be used to better understand mathematical descriptions of real cascades.

\begin{figure}[h]
    \begin{center}
    \includegraphics[width=1\textwidth]{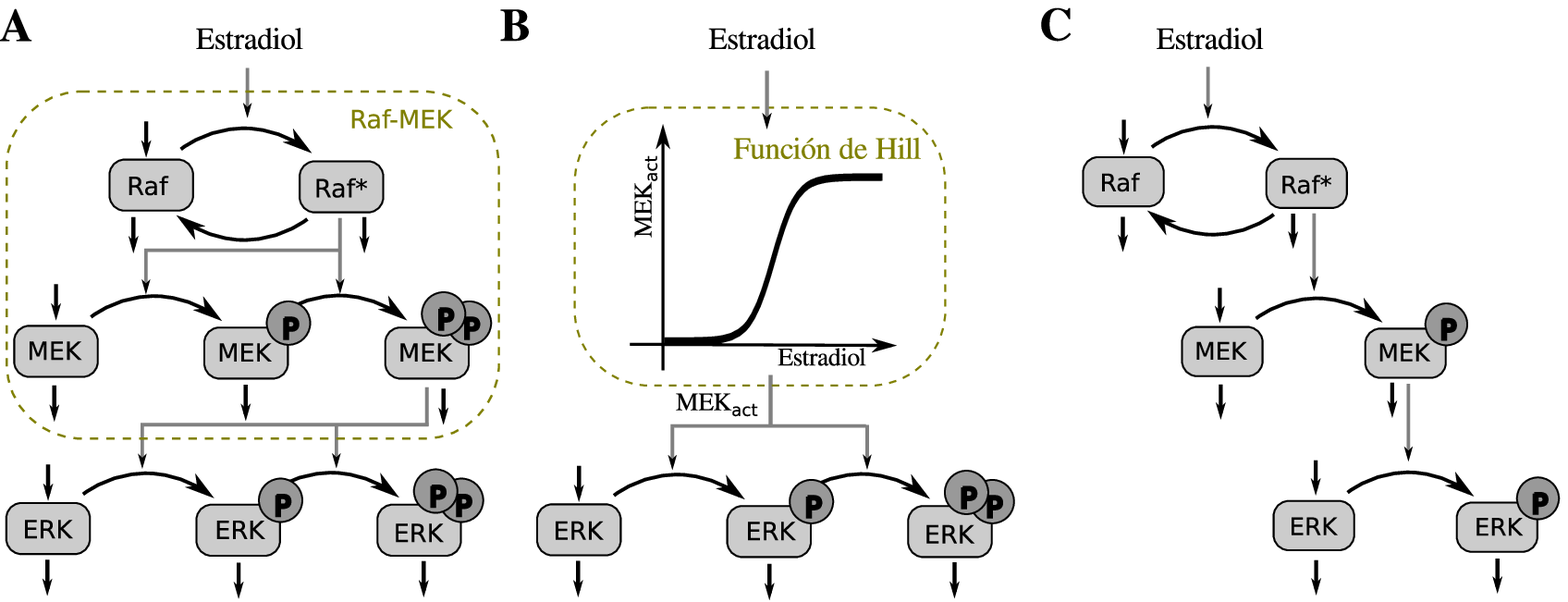}
    \caption{{\bf O’Shaughnessy et al. cascades scheme.}\\
  MAPK cascade with a dual-step phosphorylation (A) and MAPK cascade with Raf-MEK system replaced by a Hill Function (B), and a MAPK cascade with a single-step phosphorylation (C). In each case, Estradiol is the input and the most phosphorylated state of ERK is the output. In the cascade shown in (B) MEKpp in all its forms is the input to ERK layer and output of the Raf-MEK system. The Hill function parameters were worked out by fitting the active MEK dose-response by a Hill function} \label{fig_sarkar_model}
    \end{center}
\end{figure}

The three analyzed mathematical models are shown in Fig.\ref{fig_sarkar_model}. A three-tier dual-step phosphorylation cascade, a phenomenological scheme that lumps together the Raf and MEK tiers, and finally a three-tier single-step phosphorylation cascade are shown in panels (A), (B) and (C) respectively (for model details see appendix \ref{S2_text}).

\subsubsection*{Ultrasensitivity in the  dual-step phosphorylation model} 

A sketch of this model is shown in Fig.\ref{fig_sarkar_model}A. In our analysis we define the output of a module and the input to the next one as the total active form of a species, including complexes with the next layer's substrates. However we exclude complexes formed by same layer components (such as a complex between the phosphorylated kinase and its phosphatase), since these species are 'internal' to each module. By doing this, we are able to consistently identify layers with modules (the same input/output definition was used by Ventura {\em et al} \cite{ventura2008hidden}).

The analysis of $F^{non-seq}$,i.e. the mathematical composition of the  response functions of the isolated modules, allows us to assess for the effects of module cascading. We observed that the module coupling induced the existence of non-trivial HIWRs that ended up in a system-level ultrasensitivity of $n_H^{non-seq}=3.91$. This value was lower than the mere product of each module's Hill coefficient ($n_1^{is}n_2^{is}n_3^{is}=5.02$), making the cascade to have a sub-multiplicative behavior.

\begin{figure}[h]
     \begin{center}
     \includegraphics[width=1\textwidth]{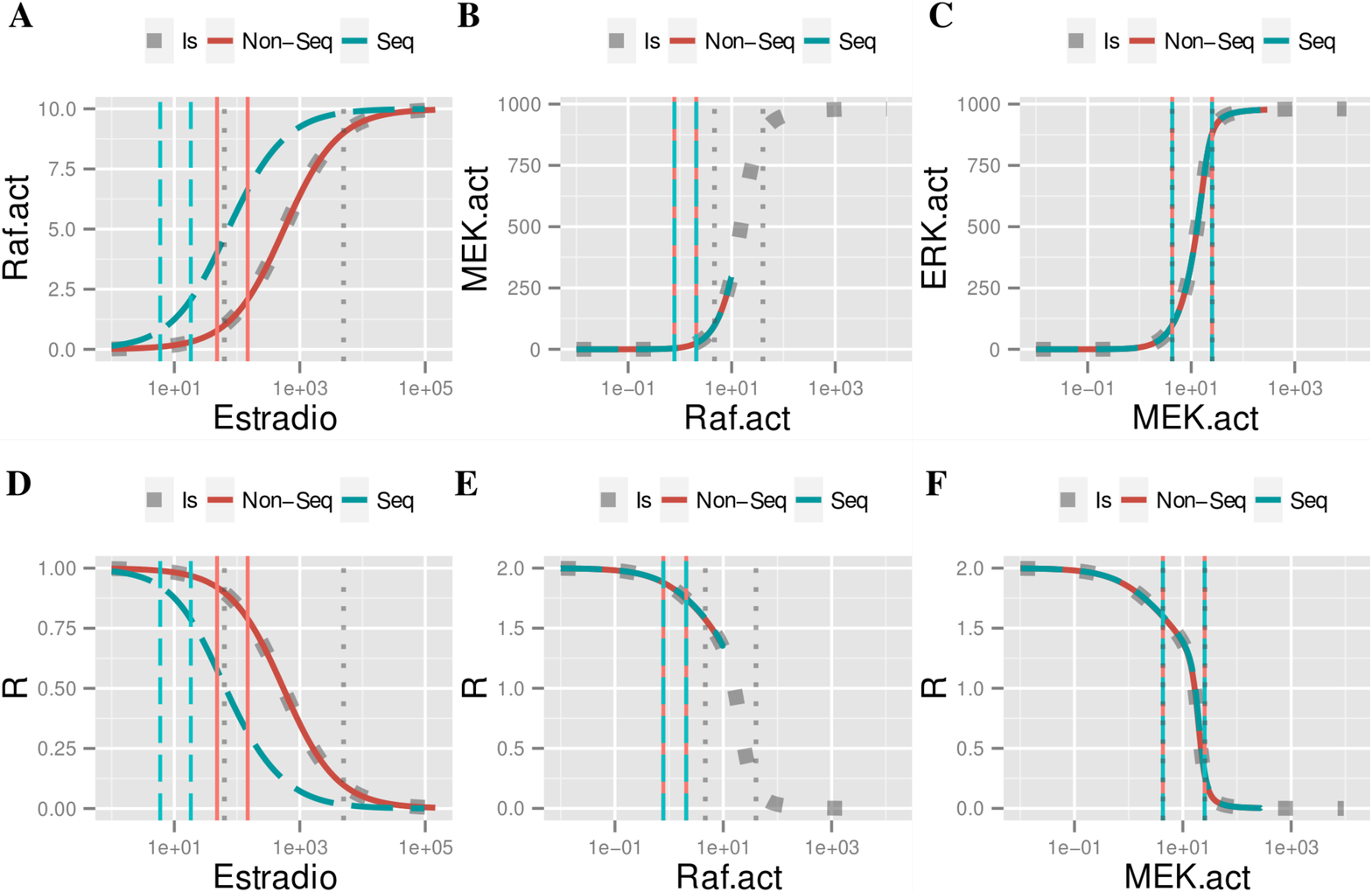}
     \caption{ \textbf{Dose-response analysis for the dual step phosphorylation model.} \\
     Transfer functions for the three layers of the MAPK cascade are depicted in panels (A-C). For each tier the isolated module transfer function (Is),  the transfer function that this module is actually sustaining in the cascade according to a mechanistic implementation of the model (Seq) and the functional form of the mathematical composition of isolated response functions (Non-Seq), are shown. The corresponding response coefficient curves are shown in panels (D-F). }\label{figSequestration}
     \end{center}
 \end{figure}

Notably, it is easy to show that sequestration is not affecting the ultrasensitivity of this system. To understand why this is the case, we refer to Fig. \ref{figSequestration}, which shows the Estradiol-act:Raf, act:Raf-act:MEK, and act:MEK-act:ERK response functions for the dual-step phosphorylation model,  in panels A, B and C respectively. Transfer functions obtained for isolated modules, the mathematical composition of corresponding isolated response functions, and the response function obtained from the mechanistic model are depicted using a dotted gray line, a continuous red line, and a dashed turquoise line respectively. The corresponding response coefficient curves are shown in panels D-F. Blue dashed vertical lines show the $X10_i$ and $X90_i$ values of each layer (i.e. mechanistic scheme), while red solid vertical lines depict layer's $X10_i$ and $X90_i$ levels associated to the composition of response curves of each module (i.e. $F^{non-seq}$). 

It can be appreciated from panels (B) and (C) of Fig.\ref{figSequestration}, that sequestration effects were actually negligible for the MAPKK and MAPK layers, given the overlap observed for the corresponding $F^{non-seq}$ and $F^{seq}$ response functions. Only for the MAPKKK layer, sequestration effects produce a shift between these curves (see Fig.\ref{figSequestration} A). 
However, it is worth noting that the corresponding Hill working ranges changed so that the effective ultrasensitive coefficients remained unchanged (i.e. $\nu_{Estradiol-Raf.act}^{Seq}=\nu_{Estradiol-Raf.act}^{Non-Seq}$). In this way the resulting overall ultrasensitivity did not get affected at all: for both implementations we get $n_H^{Seq}=n_H^{Non-Seq}=3.91$. 

Hence, we conclude that in this particular mathematical model, even though sequestration effects existed, the overall sub-multiplicative behavior was only due to a resetting of the Hill input’s working range for the first and second levels of the cascade.

Similar conclusions can be drawn for the single step phosphorylation cascade (data not shown).
 
\subsubsection*{The Raf-MEK lumped model}

In order to support the hypothesis that a cascading effect contributed to the system ultrasensitivity, O’Shaughnessy \textit{et} al. \cite{o2011tunable} analyzed the MAPK cascade with Raf-MEK levels replaced by a Hill function (Fig. \ref{fig_sarkar_model} panels A-B). They observed that this replacement produced a decrease in the cascade's ultrasensitivity, and proposed that the presence of intermediate species (MEKpp complexes) were the origin of this ultrasensitivity.

In order to understand the ultimate origin of this behavior we relied on the obtained relationship between local and global ultrasensitivity descriptors . Like O’Shaughnessy \textit{et al.} \cite{o2011tunable}, we observed a reduction of the cascade's ultrasensitivity when the Raf and MEK levels were aggregated into a Hill function, from $n_H=3.91$ to $n_H^{fhill}=2.7$ respectively. Taking into account Eq.\ref{nh_vs_mu}, Hill coefficients can be written as a function of the effective response coefficients. For the two analyzed cascades:
\begin{eqnarray}
 n_H & = &  2 \,  \nu_{Raf} \, \nu_{MEK} \, \nu_{ERK} = 2 \, \nu_{Raf-MEK} \, \nu_{ERK} \\
 n_H^{fhill} & = & 2 \, \nu_{Raf-MEK}^{fhill} \, \nu_{ERK}^{fhill}  \label{rafmek_eq}
\end{eqnarray}
 
Given that the Hill approximating function fits rather well the Estradiol-MEK curve, the HIWR of the ERK tier remains the same in both cases, and thus $\nu_{ERK}=\nu_{ERK}^{fhill}$. Then, $n_H>n_H^{fhill}$ means that $\nu_{Raf-MEK} >\nu_{Raf-MEK}^{fhill}$.

Hence, the observed overall ultrasensitivity reduction in the lumped model is due to a reduction of the effective response coefficient of the Raf-MEK Hill approximating function, $\nu_{Raf-MEK}^{fhill}$, with respect to the effective response coefficient  $\nu_{Raf-MEK}$, associated to the Estradiol-MEK response curve in the original model (dashed boxes in Fig. \ref{fig_sarkar_model}B and A respectively). 
We calculated the effective response coefficient of each layer, obtaining a $\nu_{Raf-MEK}=1.58$ and a $\nu_{Hill\,func}=1.09$ which is consistent with our expectations. The combined Raf-MEK layers have a Hill coefficient of $n_H^{Raf-MEK}=n_H^{fHill}=1.14$. This means that while the Raf-MEK system is contributing to the original cascade with an ultrasenstivity higher that its Hill coefficient, this is not the case for the lumped model.

\begin{figure}[h]
    \begin{center}
    \includegraphics[width=0.4\textwidth]{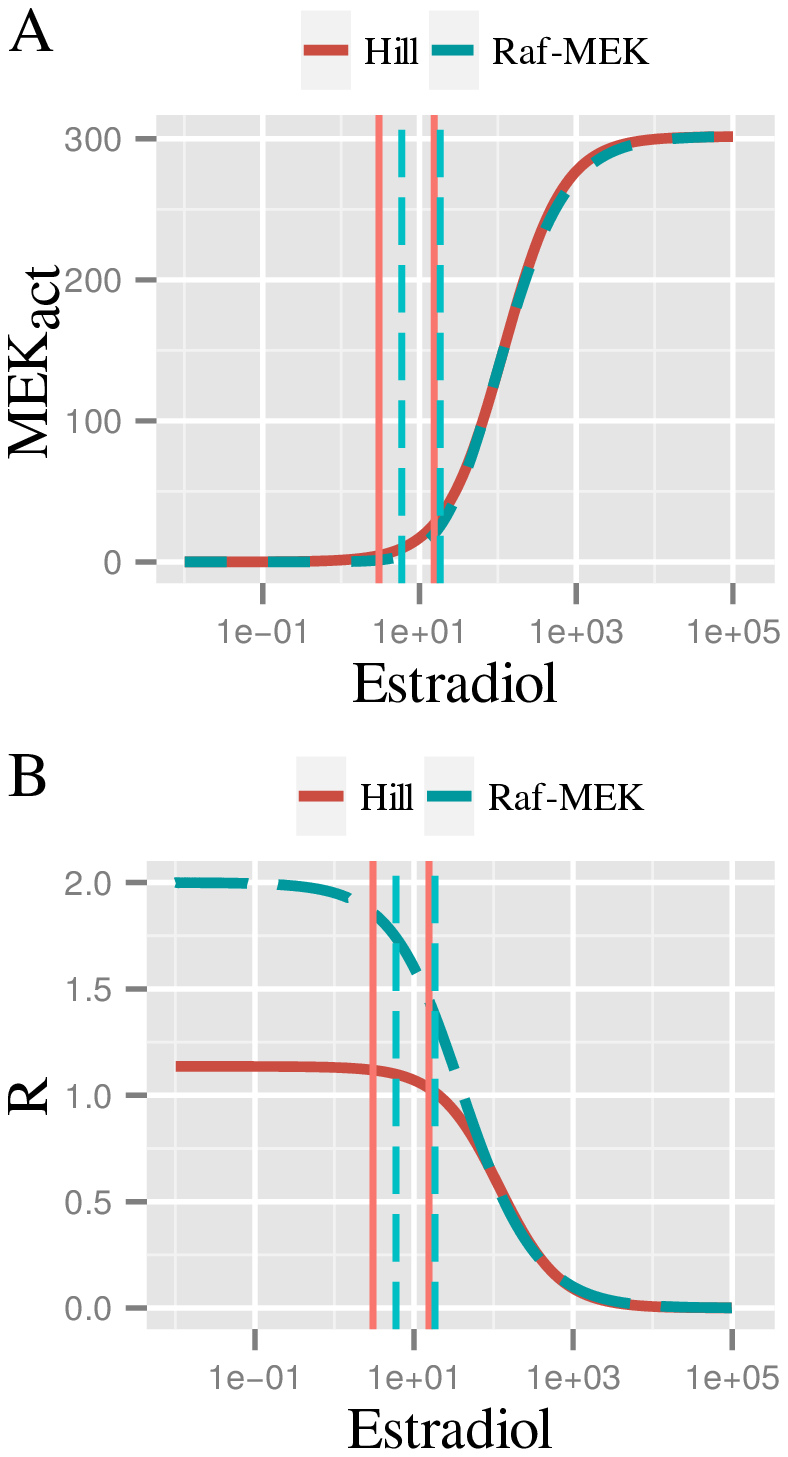}
    \caption{{\bf  Fitting by a Hill function may neglect relevant behaviors.}\\
 Dose-response curve of active MEK in O’Shaughnessy model compared with its fit by a Hill function (A). Respective response coefficient (B). It can be seen that as the dose-response of active MEK and the Hill function appear to be similar, there are strong dissimilarities in their local ultrasensitivity}\label{fig_MAPK_cascades_R}
    \end{center}    
\end{figure}

The cause of this behavior can be understood looking at Fig. \ref{fig_MAPK_cascades_R}. Even though the dose-response of active MEK and the Hill approximating function appear to be identical, there are strong dissimilarities in their local ultrasensitivity behavior. This is particularly true for low input values, where the Hill's input working range is located. In this region, the active MEK curve presents local ultrasensitivity values larger than the Hill function counterpart, thus the replacement by a Hill function produces a reduction in the corresponding Hill coefficient. In this way, despite the high-quality of the fitting adjustment (Residual Standard Error=2.6), the Hill function approximation introduced significant alterations in the system’s ultrasensitivity as a technical glitch.

This is a remarkable result as it states that a well approximating function from the point of view of standard minimization procedures, might non-trivially impinge on qualitative conclusions about the system behavior.

\subsubsection*{The single-step phosphorylation model}

In order to probe the origin of the ultrasensitivity observed in the original cascade (sketched in Fig \ref{fig_sarkar_model}A), O’Shaughnessy et al. constructed an auxiliary model in which dual-step phosphorylation tiers where replaced by single-step phosphorylation layers. Because in this new setting the cascade is not subject to multiple activation processes, competitive inhibition nor zero-order ultrasensitivity (due to the absence of phosphatases), they claim that there is no other ultrasensitivity source than the kinase-cascading architecture itself. Thus, they propose that the ultrasensitivity observed in this cascade is due to a \textit{de-novo} ultrasensitivity generation. 

\begin{figure}[h]
    \begin{center}
    \includegraphics[width=0.8\textwidth]{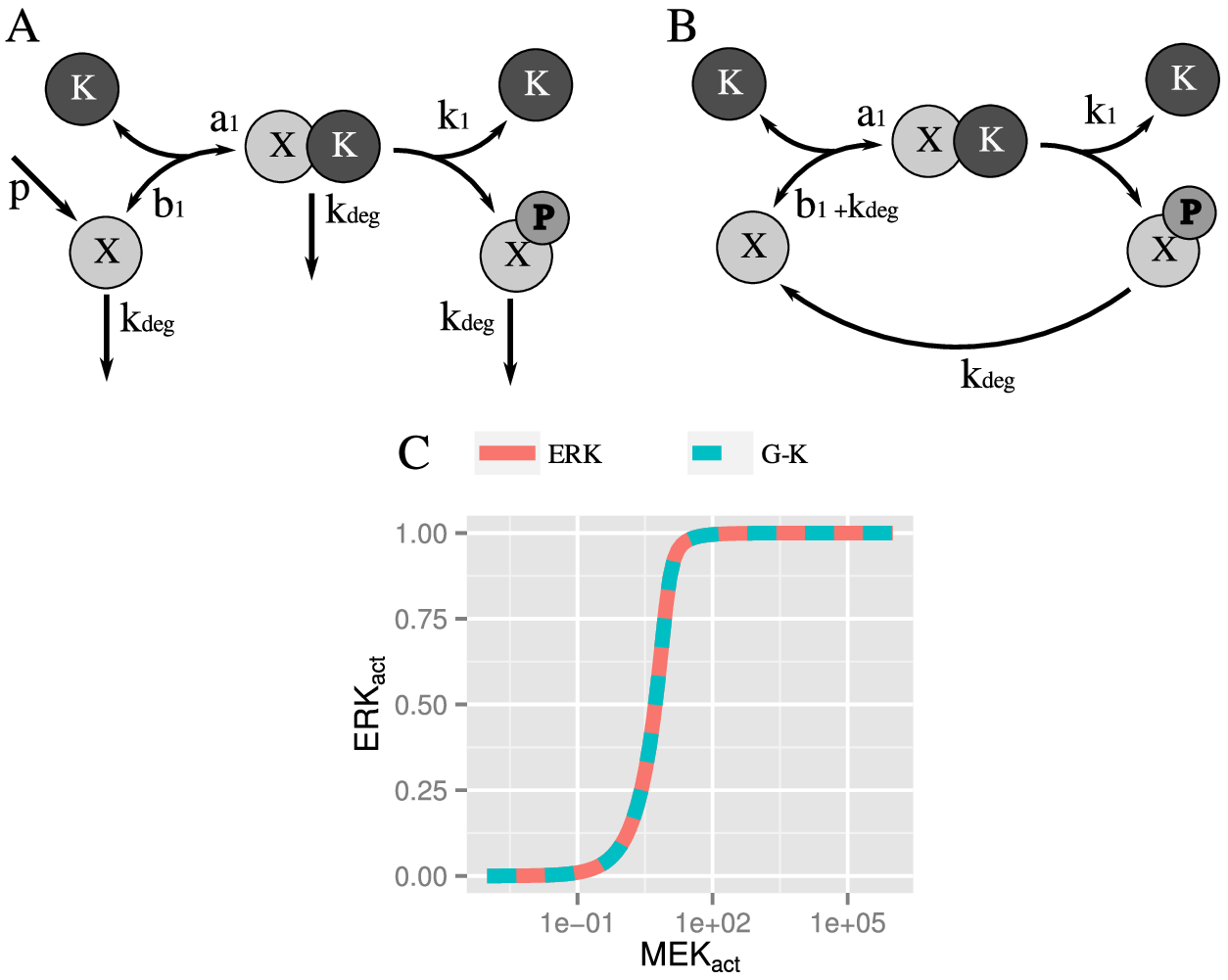}
    \caption{{\bf  Equivalence between a single-step layer in O’Shaughnessy model and a covalent cycle.}\\
  O’Shaughnessy et al. single-Step layer scheme (A) and equivalent covalent cycle scheme (B). Steady state transfer function of ERK layer in isolation of the O’Shaughnessy single-step cascade in blue dashed line (C), compared to a centered Goldbeter-Function with equivalent parameters in red solid line ($K_1=0.04$ and $K_2=1000$, see appendix \ref{S1_text} )}\label{fig_covalent_cycles}
    \end{center}
\end{figure}

Despite of this claim, we observed that when MEK and ERK modules are considered in isolation, they did present an ultrasensitive behavior ($n^{MEK}=1.54$ and $n^{ERK}=1.76$). Synthesis and degradation happened to be the key factors to understand the origin of the ultrasensitivity. This layers (scheme in Fig \ref{fig_covalent_cycles}A) are in fact mathematically analogous to a covalent cycle (scheme in Fig \ref{fig_covalent_cycles}B) because there is an implicit channel from the activated protein towards its inactive form via the degradation of the active protein and the production of the inactive form. Given that degradation is a linear reaction with respect to the amount of activated protein, its mathematical description is equivalent to a dephosphorylation reaction operating in a first order regime. Equivalently it can be considered as a limit case where the complex formed by the active protein and phosphatase instantly disassembles (i.e. $K_2 \rightarrow \infty$)

Thus, the one-step system depicted in Fig \ref{fig_covalent_cycles} could in fact be described by a Goldbeter-Koshland (G-K) \cite{goldbeter1981amplified} function with

\begin{equation}
 K_1 = \frac{K_{deg}+b_1+k_1}{X_T a_1} \, \, \, and  \, \, \, K_2 \gg 1
\end{equation}

We plotted in Fig \ref{fig_covalent_cycles}C the steady state transfer function of the ERK module in isolation and the corresponding centered G-K function (see appendix \ref{S1_text}). A clear agreement between both functions can be appreciated. In the light of these results we conclude that the single-step cascade’s ultrasensitivity did not come from a cascading effect but from a ``hidden'' first-order ultrasensitivity process in the MEK and ERK layer.

\section*{Discussion}
The study of signal transmission and information processing inside the cell has been, and still is, an active field of research. In particular, the analysis of signaling cascades has received a lot of attention as they are well-conserved motifs that can be found in many cell fate decision systems. The aim of this paper is to propose a framework to characterize and better understand mathematical models used to study real biological systems. For a given mathematical model, the  methodology we described allowed us to disentangle the origin of the predicted ultrasensitivity behavior in terms of Hill's input working range resetting and/or sequestration effects acting on the modular cascade architecture of interest. In this respect, even though we have not dealt with the general and important problem of resolving the working principles acting on a given real cascade, we did provide a useful tool to modelers in order to better understand and perform educated choices between modeling alternatives. 

It is also worth noting that dynamical features of signal transduction systems might play an important role on the system-level displayed behavior. In order to analyze signaling cascades whenever this happens, one should not only deal with the coupling of modular input-output response functions but also with their characteristic time-scales. Despite of this a steady state analysis, such as the one presented here, still offers useful information and remains a sensible approximation whenever there is no effective time-scale separation and/or upstream modules happens to evolve faster than downstream ones.

In this work we have found a mathematical expression (equation \ref{nh_vs_R}) that linked local and global ultrasensitivity descriptors in a fairly simple way. Moreover we have provided a general result to handle the case of a linear arrangement of an arbitrary number of such modules (equation \ref{nh_vs_mu}).
The value of the obtained expression  resides in the fact that not only it captured previous results, like Ferrell’s inequality, but also in that it threw light on the mechanisms involved in ultrasensitivity generation. For instance, the existence of supramultiplicative behaviors in signaling cascades have been reported by several authors \cite{racz2008sensitivity,o2011tunable} but in many cases the ultimate origin of supramultiplicativity remained elusive. Our framework naturally suggested a general scenario where supramultiplicative behavior could take place. This could occur when, for a given module, the corresponding \textit{Hill’s input working range} was located in an input region with local ultrasensitivities higher than the global ultrasensitivity of the respective dose-response curve.

Notably, within the proposed analysis framework, we could decompose the overall global ultrasensitivity in terms of a product of single layer effective response coefficients. These new parameters were calculated as local-sensitivity values averaged over meaningful working ranges (here called \textit{Hill’s input working ranges}), which permitted to assess the effective contribution of each module to the system’s overall ultrasensitivity. Of course, the reason why we could state an exact general equation for a system-level feature in terms of individual modular information was that in fact system-level information was implicitely used in the definition of  \textit{Hill’s input working ranges} that entered equation \ref{nh_vs_mu}. The specific coupling between ultrasensitive curves set the corresponding \textit{Hill’s input working ranges}, thus determining the effective contribution of each module to the cascade’s ultrasensitivity. This process, which we called \textit{Hill’s input working range setting}, has already been noticed by several authors \cite{altszyler2014impact,ferrell1997responses,brown1997protein,bluthgen2001map,bluthgen2003robust,racz2008sensitivity,tsvetanova2016gpcr}, but as far as we know this is the first time that a mathematical framework, like the one we present here, has been proposed for it.

We used our methodology to revisit the different mathematical models considered by O’Shaughnessy et al. to analyzed their tunable synthetic MAPK system \cite{o2011tunable}, and we were able to bring a new perspective to the conclusions that could be drawn from such mathematical constructs.

For instance, we proved that sequestration effects played no role in the observed system ultrasensitivity for the dual-step and single-step phosphorylation models. We also were able to analyze the auxiliary model in which the Raf and MEK layers were replaced by a Hill function that is coupled to the ERK layer. In this case, even though the original Estradiol-MEK input-output response curve could be fairly well fitted and global ultrasensitivity features were rather well captured, the mere replacement by a Hill function produced a strong decrease in the systems ultrasensitivity. We found that the functional form of the Hill function failed to reproduce original local ultrasensitivity features that were in fact the ones that, due to the particular \textit{Hill working range setting} acting in this case, were responsible for the overall systems ultrasensitivity behavior. The analyzed case was particularly relevant, as provided an illustrative example that warned against possible technical glitches that could arise as a consequence of the inclusion of approximating functions in MAPK models.

\section*{Conclusions}

In the present article we provided a framework to characterize mathematical models used to describe real biological systems of ultrasensitive character. We presented a mathematical link between global and local ultrasensitivity estimators for a sigmoidal unit and generalized these results for a cascade of such units. Using the introduced concept of {\em Hill input's working range}, the overall system's ultrasensitivity could be defined in terms of  effective contributions of each cascade tier. Moreover, we were able to explain the origin of the ultrasensitivity in a given mathematical model in terms of Hill's input working range resetting and/or sequestration effects. 

Our framework may help to understand the origin of ultrasensitivity in general multilayer structures, and in this sense it could be useful in the design of synthetic systems \cite{zechner2016molecular, kittleson2012successes, ang2013tuning}. 
For instance, given that the right working range setting (targeting the region of maximal local ultrasensitivity of a given unit in a cascade) is a key factor in producing high overall ultrasensitivity, our methodology can be used to guide the tuning of a single module's features, as well as its coupling with other units forming a cascade, in order to control the system's ultrasensitivity. 

\newpage
\appendix

\section{Appendix: Goldbeter-Koshland function} \label{S1_text}
\begin{figure}[h]
    \begin{center}
    \includegraphics[width=0.5\textwidth]{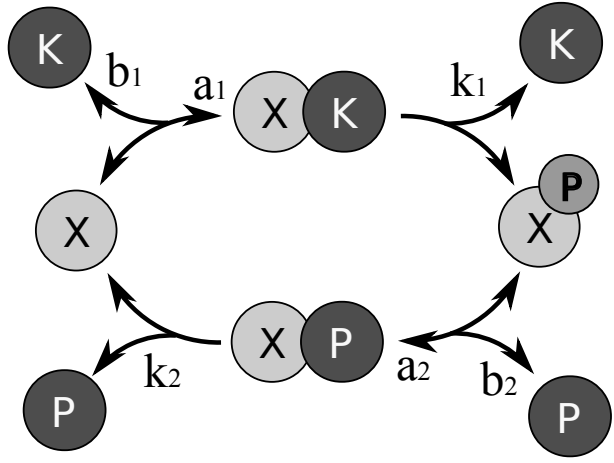}
    \caption{{\bf Covalent cycle wiring diagram.}   }\label{fig_cov_cycl}
    \end{center}
    
\end{figure}

The Goldbeter-Koshland function \cite{goldbeter1981amplified} is used to describe the steady state concentration of a protein affected by phosphorylation/de-phosphorylation modifications (schematized in Fig \ref{fig_cov_cycl}). 

It's transfer function is given by the expression:

\begin{equation}
 G(x,K_1,K_2) = \frac{(x-1)-K_2 \Big( x+ \frac{K_1}{K_2}  \Big) + \Big( x-1-K_2 \Big( x+ \frac{K_1}{K_2}  \Big) -4 K_2 (x-1) x \Big)^{1/2}}{2(x-1)}
 \label{gkFunction}
\end{equation}
with 
\begin{equation}
x=\frac{k_1 K_T}{k_2 P_T}  \, \, \, , \, \, \,  K_1 = \frac{b_1+k_1}{X_T a_1} \, \, \, and  \, \, \,  K_2 = \frac{b_2+k_2}{X_T a_2}
\end{equation}
where $X_T$, $K_T$ and $P_T$ are the total concentration of proteins X, K and P, respectively. $K_1$ and $K_2$ are the phosphorylation and de-phosphorylation Michaelis constants divided by $X_T$, respectively.

In order to center the G-K function, we multiply the independent variable for a scale factor $\alpha$, $G(\alpha, x, K_1, K_2)$, where $\alpha$ was set in order to make the EC50 of G-K function coincides with the desired EC50 value.

\newpage
\section{Appendix: O’Shaughnessy \textit{et al.} model description}\label{S2_text}

For the sake of completeness we included in Fig.\ref{reactions} the description of O’Shaughnessy's ODEs models (see supplementary tables S5, S7, S8 of O’Shaughnessy's \textit{et al.} paper \cite{o2011tunable}). 

\begin{figure}[h]
    \begin{center}
    \includegraphics[width=0.8\textwidth]{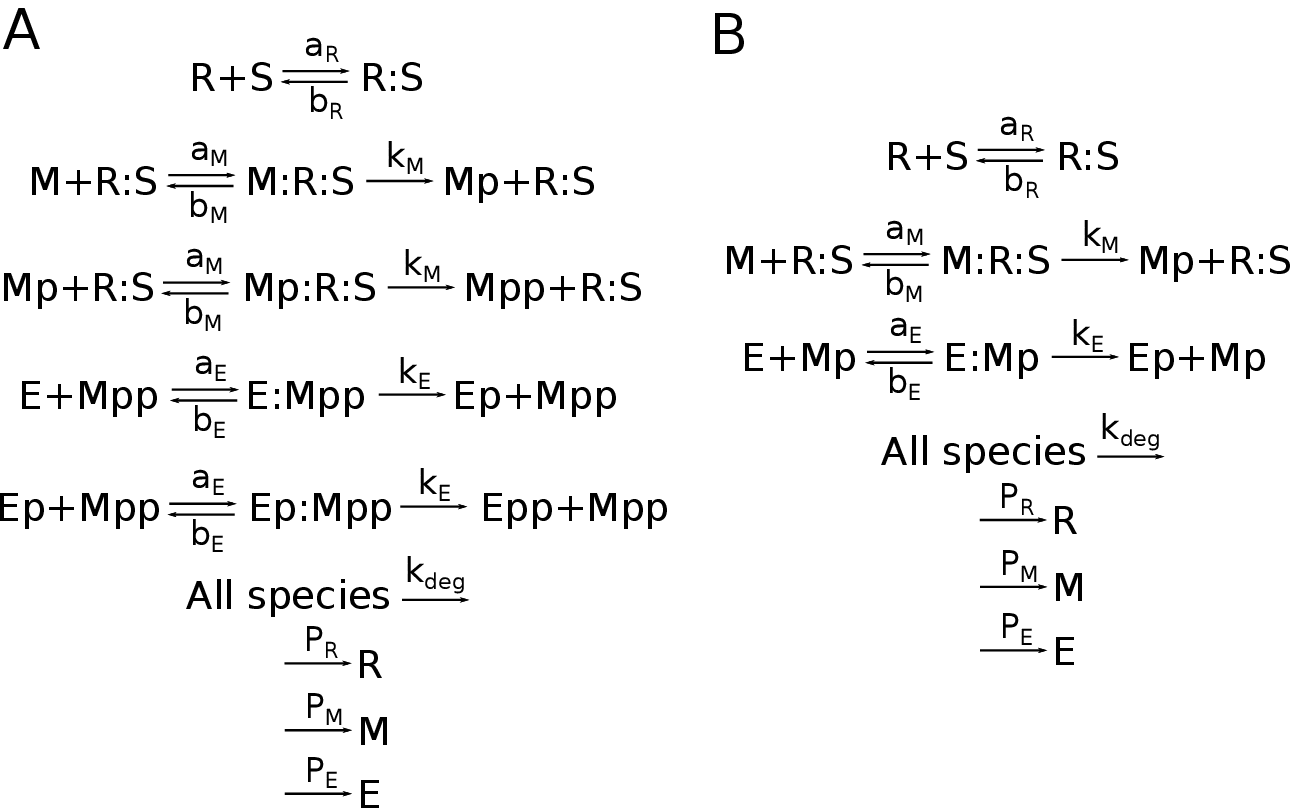}
    \caption{{\bf  O’Shaughnessy \textit{et al.} models reactions.}\\
    A) Dual-step model reactions and B) single-step model reactions. }
    \label{reactions}
    \end{center}
\end{figure}
\subsection*{Parameters:}
$a_R=0.9 \frac{1}{\mu M.s}$; $a_M=5 \frac{1}{\mu M.s}$; $a_E=15 \frac{1}{\mu M.s}$; $k_M=k_E=0.1 \frac{1}{\mu s}$; $b_R=b_M=b_E= 0.5\frac{1}{s}$; $deg=0.001\frac{1}{s}$; $P_R=0.01$; $P_M=P_E=1$

\subsection*{Initial conditions:}
$R=10nM$; $M=1000nM$;$E=1000nM$; and all other species start with zero concentration. It is worth noting that the total concentration of R, M and E, in all their forms will be held constant given the balance of protein production and degradation. 


\newpage
\section{Appendix:}
\begin{figure}[h]
    \begin{center}
    \includegraphics[width=1\textwidth]{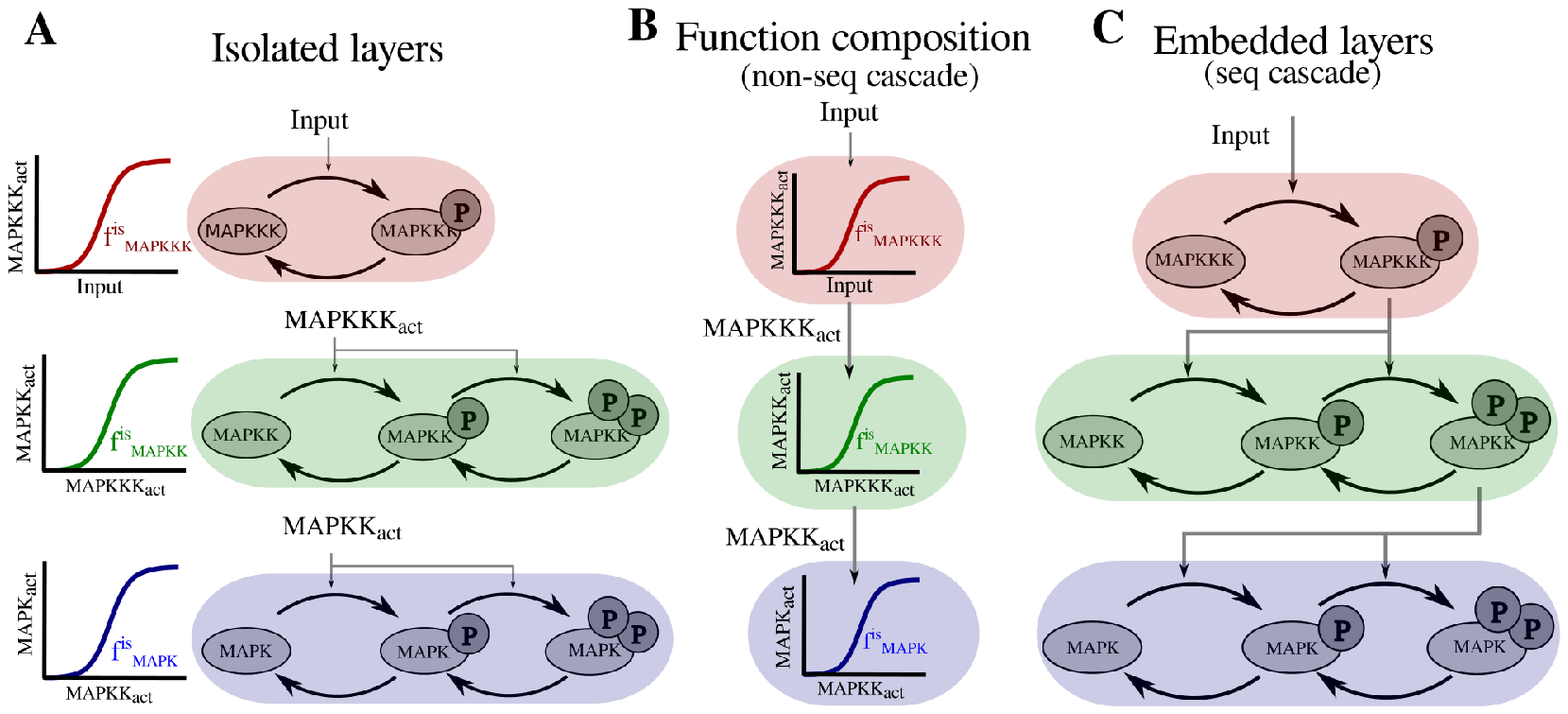}
    \caption{{\bf Modular and system representation of a MAP kinase cascade.}\\
Each layer in isolation is composed by single or multiple covalent cycles, which dose-response curves can be ultrasensitive by zero-order mechanisms and/or multi-activation processes (A). The cascade transfer function, in a scenario in which sequestration is not taken into account ($F^{non-seq}$) can be obtain by the mathematical composition of each module’s transfer functions $f^{is}_i$ acting in isolation (B). When the sequestration effect is taken into account, the layers embedded in the MAP kinase cascade may have a different dose-response curve from the isolated case (C). }\label{S1_Fig}
    \end{center}
\end{figure}
\newpage

\bibliographystyle{unsrt}
\bibliography{Altszyler_ultrasensitivity_arxiv_2017}{}

\end{document}